\documentclass[aps,prl,amssymb,showpacs,floatfix,twocolumn]{revtex4}
\usepackage{amsmath,amsfonts,graphics,epsfig,amssymb}
\begin{document}
\title{%
Constraints on the flux of primary cosmic-ray photons at energies
$E>10^{18}$~eV\\
  from Yakutsk muon data
}
\author{A.V.~Glushkov}
\author{I.T.~Makarov}
\author{M.I.~Pravdin}
\author{I.E. Sleptsov
(Yakutsk EAS Array)}
\affiliation{
Yu.G.~Shafer Institute of Cosmophysical Research and
  Aeronomy,
  Yakutsk 677980, Russia
}
\author{D.S.~Gorbunov}
\author{G.I.~Rubtsov}
\author{S.V.~Troitsky}
\affiliation{Institute for Nuclear Research of RAS, 60th October
Anniversary Prospect 7a, Moscow 117312, Russia}
\date{v.2: September 24, 2009}
\begin{center}
\begin{abstract}
Comparing the signals measured by the surface and underground
scintillator detectors of the Yakutsk Extensive Air Shower Array, we place
upper limits on the integral flux and the fraction of primary cosmic-ray
photons with energies $E>10^{18}$~eV,
$E>2\times 10^{18}$~eV and $E>4\times 10^{18}$~eV. The large collected
statistics of the showers measured by large-area muon detectors
provides a sensitivity to photon fractions $<10^{-2}$,
thus achieving precision
previously unreachable at ultra-high energies.
\end{abstract}
\end{center}
\maketitle


\subsection{Introduction}
Ultra-high-energy (UHE) cosmic-ray (CR) photons are
produced by energetic protons and nuclei
in their interactions both at acceleration sites and
along their trajectories towards the Earth~\cite{Ginzburg:1990sk}.
Both protons and
heavier nuclei with energies $E \sim 10^{20}$~eV
interact with cosmic background radiations, especially with
cosmic microwave background (CMB) and infrared
background (IRB) radiation.
The processes involved in these interactions are however very
different. Interactions of a {\it proton} at $E \gtrsim 7 \times
10^{19}$~eV with CMB photons lead to efficient pion production
\cite{Greisen:1966jv, Zatsepin:1966jv}.
Further decays of neutral pions produced in these
interactions lead to a secondary photon flux at energies $E\gtrsim
10^{18}$~eV (so-called Greisen-Zatsepin-Kuzmin photons) \cite{Lee:1996fp, Gelmini:2005wu}. On
the other hand, the dominant interaction channel for {\it heavier nuclei}
is their photodisintegration on IRB photons;
the secondary
photon flux is much smaller in this case \cite{Gelmini:2007jy}. Therefore,
the photon flux at  $E\gtrsim 10^{18}$~eV may provide an independent test
of the chemical composition of CRs at $E\sim (10^{19}\dots
10^{20})$~eV which is, at present, largely
uncertain~\cite{HiRes:comp, HiRes:comp:fluct, PAO:comp, bimodal}.

On the other hand, the study of UHE photons is a powerful tool for
constraining new-physics models. One example is provided by
models with superheavy dark-matter (SHDM) particles (e.g.\
\cite{Berezinsky:1997hy});
a substantial fraction of the SHDM decay products
are photons.
Another class of exotic relics to be searched for with CRs is
topological defects
\cite{Hill:1986mn, Bhattacharjee:1998qc};
UHE photons were
suggested
\cite{Berezinsky:1998ft} as their signature.
With the help of UHE photons, one may also constrain astrophysical models
of the CR origin which involve new physics at the propagation
stage. In particular,
both the spectrum and the chemical composition of CRs are
changed in models with violation of the Lorentz invariance
\cite{Coleman:1998ti}.
The photon fraction at the highest energies is sensitive to parameters
violating Lorentz invariance, and upper limits on the former severely
constrain the latter \cite{Galaverni:2007tq}. Finally, photons with
energies above $\sim 10^{18}$~eV might be responsible for CR
events correlated with BL Lac type objects on  the angular scale
significantly smaller than the expected deflection of protons in cosmic
magnetic fields and thus suggesting neutral
primaries~\cite{Gorbunov:2004bs, Abbasi:2005qy} (see Ref.~\cite{axion} for
a particular mechanism).

In this paper, we present the analysis of extensive air showers
observed by the Yakutsk extensive-air-shower array,
which yields the strongest
limits on the photon flux and the photon fraction in CRs at
energies $E>10^{18}$~eV, $E>2\times 10^{18}$~eV and $E>4\times
10^{18}$~eV. These limits enter the region interesting both for
highest-energy astrophysics and tests of extragalactic backgrounds as well
as for searches of new physics.
To obtain better quantitative constraints it would be helpful to use the
results of other experiments jointly with ours; it
will be done elsewhere.


\subsection{Method}
The key idea of our method is the event-by-event comparison
of the observed muon densities in air showers with those in simulated
gamma-ray induced showers which have the same signal density as
measured by the surface scintillator array and have the same arrival
direction as the observed ones.  The method is described in detail in
Ref.~\cite{composition}; it has been previously applied to Yakutsk and
AGASA muon data at the highest energies~\cite{AYgamma, Ygamma}.
Similar statistical methods have been used to constrain primary photon
content from the data collected by the fluorescent detector of the
Pierre Auger Observatory~\cite{PAO:FDgammaOld} (see also
Ref.~\cite{Homola}).  One of the advantages of the method is its
independence both from the energy-reconstruction procedure used by the
experiment and from the Monte-Carlo simulation of hadronic air
showers: we use simulated gamma-ray induced showers which are mostly
electromagnetic and are therefore well understood and we select the
simulated showers by the observable signal density in the surface
scintillator array and not by the energy (effectively estimating the
energy of each event in the assumption of a photon primary).

\begin{figure}
\centering \includegraphics[width=0.95\columnwidth]{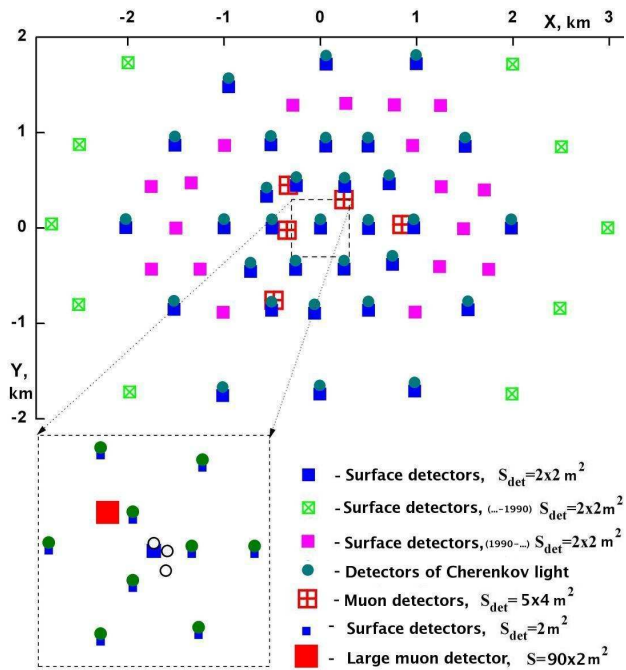}
\caption{
\label{fig:array}
A sketch of the Yakutsk array. Each surface detector
consists of two 2\,m$^2$ scintillators, while each muon detector consists
of five 4\,m$^2$ scintillators placed underground (with shielding
equivalent to about 2\,m thick concrete wall). The signal of the large
muon detector is not used in this study as well as signal of the small
central subarray outlined separately. }
\end{figure}

The Yakutsk extensive-air-shower array (Yakutsk, Russia)
has been observing
UHECR events since 1973,  with detectors in various
configurations~\cite{Egorova:2004cm, PravdinQ2004, Glushkov_ICRC2003}
covering an area from 10\,km$^2$ to 20\,km$^2$ in different operation periods.
For the data set used in this work, the surface array comprised 49 (before
1990, 41) detectors, where each detector consisted of two 2~m$^2$
scintillation counters. It is equipped, since 1982, with five muon
detectors of 20~m$^2$ area each with threshold energy 1~GeV for vertical
muons~\cite{muonLDF}. A sketch of the Yakutsk array is presented in
Fig.~\ref{fig:array}.

At present, it is the only installation in
the world which is equipped with muon detectors and capable of studying
CRs with energies above $10^{18}$~eV.
The surface scintillator detector signal density at 600~m from the shower
axis, $S(600)$, together with the shower geometry, is obtained from a
joint fit of the lateral distribution function and the shower front
arrival times~\cite{Yellow}. For the events we use, the angular resolution
is $\approx 5^\circ$ and the mean $S(600)$ resolution is $\approx 17\%$.
The energy of a primary particle is estimated in the Yakutsk
experiment from $S(600)$ and the zenith angle as described in
Refs.~\cite{Glushkov_ICRC2003,YakEnergy}, by making use of the
experimental
calibration to the atmospheric Cerenkov light \cite{YakCalib}
and of the attenuation curve
determined from data by means of the constant intensity cuts method
suggested in Ref.~\cite{CIC:Hillas} and used also by the Haverah
Park~\cite{CIC:Haverah}, AGASA~\cite{CIC:AGASA} and Auger~\cite{CIC:Auger}
experiments. This reconstructed energy $E_{\rm est}$ may differ from the
true primary energy $E$ both due to natural fluctuations and due to
possible systematic effects. These latter effects depend on the primary
particle type; in particular, the difference between photons and hadrons
is significant~\cite{Nuhuil}. This difference in the energy estimation of
primary gamma rays and primary hadrons (which constitute the bulk of
observed UHECRs) forces one to use different ways of energy reconstruction
when searching for photon primaries.

For the present study, we use the sample of events satisfying the
following criteria:
(1)~the event passed the surface array trigger described in
Refs.~\cite{PravdinQ2004, Glushkov_ICRC2003};
(2)~the reconstructed core location is inside the array
boundary;
(3)~the zenith angle $\theta \le 45^\circ$;
(4)~the reconstructed energy
$E_{\rm est} \ge 10^{18}$~eV;
(5)~the reconstructed shower axis is within 300~m from
an operating muon detector.
The data set contains 1647 events observed between December 10, 1982 and
June 30, 2005 and corresponds to an effective exposure  of $7.4\times
10^{8}~{\rm km}^2\cdot {\rm s}\cdot {\rm sr}$ for $E>10^{{18}}$~eV (an
important reduction in the effective area is related to the cut~(5)).

By making use of the empirical muon lateral distribution function
\cite{muonLDF},  we calculate, for each event,
the muon density at 300~m
from the shower axis, $\rho _\mu (300)$, which we use as the composition
estimator.
We apply the event-by-event analysis following Ref.~\cite{composition}
and estimate, for each event, the probability that it has been initiated by
a primary photon.
To this end, we use
a library of $\sim 2\times 10^4$ artificial photon-induced showers with
different energies ($2\times 10^{17}$~eV$<E<2\times 10^{19}$~eV) and
arrival directions, of which we select those with
the same $S(600)$
 and zenith angle~\footnote{Since all events in the sample have
reconstructed energies below $10^{19}$~eV, we do not expect
azimuthal-angle dependence of the shower properties due to geomagnetic
cascading; therefore we require consistency between the arrival directions
of the observed and artificial showers in zenith angle only.} as the
observed event, up to reconstruction errors (a detailed description of the
method is presented in Refs.~\cite{composition, AYgamma}).

To simulate the shower library, we used CORSIKA~6.611~\cite{corsika} with
FLUKA~2006.3~\cite{fluka} as a low-energy hadronic interaction model and
EPOS~1.61~\cite{EPOS} as a high-energy model.
The difference in the expected muon density  between various models
is negligible for photon showers (we checked it explicitly for
EPOS~1.61 and QGSJET~II
\cite{QGSJET-II}). We used
thinning ($10^{-5}$) with weight limitations~\cite{Thin} to save
computational time~\footnote{This
choice of thinning introduces artificial fluctuations $\sim~5\%$ in both
$\rho_\mu$ and $S$ \cite{our-thin} which make our upper limits more
conservative.}.

\begin{figure}
  \centering
\includegraphics[width=0.95\columnwidth]{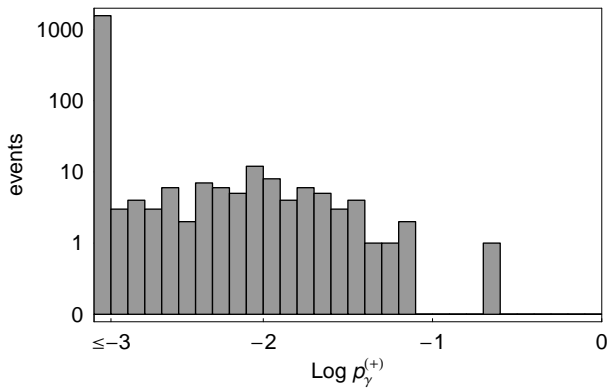}
\caption{
\label{fig:p-distr}
Distribution of probabilities $p_\gamma^{(+)}$ for the sample of real events
with the lower energy cut on energy $E_{\rm
min}=10^{18}$~eV.
}
\end{figure}

\begin{figure}
  \centering
\includegraphics[width=0.95\columnwidth]{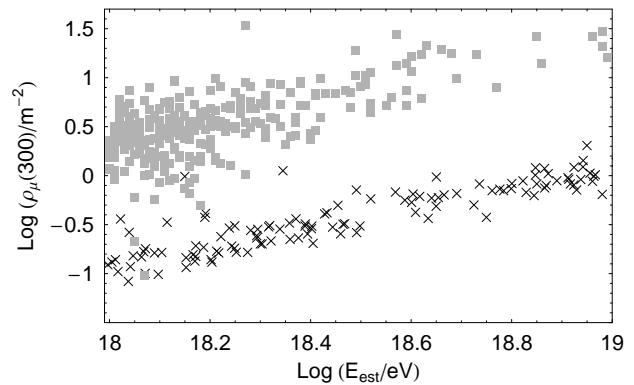}
\caption{
\label{fig:rho_mu}
Muon densities $\rho_\mu(300)$ of air showers with $30^\circ <\theta
<35^\circ$ versus primary energies $E_{\rm est}$ reconstructed by the
standard Yakutsk procedure for simulated photon-induced events (crosses)
and for real data (boxes). }
\end{figure}

For each simulated shower, we determine $S(600)$ and
$\rho_{\mu}(300)$ by making use of the GEANT simulations of the detector
\cite{YakutskGEANT}.
This enables us to select
simulated showers compatible with the observed ones
by $S(600)$:
each artificial shower gets a weight determined by the difference
in $E_{\rm est}$ from the real event~\footnote{$E_{\rm est}$ follows the
Gaussian distribution in log(energy); the standard deviation of $E_{\rm
est}$ has been determined event-by-event; on average it is $\sim$17\%
\cite{Pravdin_ICRC2005}.}. For each of the observed events, we calculate
the distribution of simulated muon densities representing photon-induced
showers compatible with the observed ones by $S(600)$ and $\theta$ in the
following way.
To take into account possible
experimental errors in the determination of the muon density, we replace
each simulated $\rho _\mu (300)$ with a Gaussian distribution
representing possible statistical errors (see formulas and discussion
in Ref.~\cite{composition}). The latter
have been estimated
for each event individually by fitting muon detector readings with
the lateral distribution function~\cite{muonLDF,composition}.
The dominant contribution to the statistical
error of $\rho _\mu (300)$ comes from the uncertainty in the determination
of the shower axis (for which we use the geometric reconstruction from the
main scintillator array). The overall uncertainty of $\rho _\mu (300)$
varies from $\sim 15\%$ to $\sim 40\%$ for individual events. The
distribution of the simulated muon densities is the weighted average of
these Gaussians. For each event in the dataset we derive, from this
distribution, the probability $p_\gamma^{(+)}$ that it has been initiated
by a primary photon of energy in the range under study ($E>E_{\rm
min}$ for $E_{\rm min}=10^{18}$~eV, $2\times 10^{18}$~eV or $4\times
10^{18}$~eV). The distribution of $p_\gamma^{(+)}$ for the
observed events is presented in  Fig.~\ref{fig:p-distr}.
Then Fig.~\ref{fig:rho_mu}
illustrates that for most events, the measured muon
densities are too high as compared to those obtained from simulations
of photon induced showers. A simple statistical
procedure~\cite{composition} allows one to determine upper limits on the
photon content from the ensemble of $p_\gamma^{(+)}$.

Below, we present limits on the fraction of gamma rays and on the
absolute gamma-ray flux. For the fraction limits, we use explicit
formulae of Ref.~\cite{composition}.  The fraction limits
depend~\cite{Ygamma} on the energy scale assumed {\it for non-photon
  primaries} which has a systematic uncertainty of
30\%~\cite{Egorova:2004cm}.
The flux limits do
not depend on the choice of hadronic interaction model used in
simulations, nor on the energy reconstruction used in the experiment; the
only assumption is that electromagnetic showers are simulated correctly.
To obtain the
limit on the flux of primary photons, we slightly modified the technical
part of the procedure of Ref.~\cite{composition}. Let $F_{\gamma}$ be the
integral flux of primary photons over a given energy range. Then we expect
to detect
$$\bar n (F_{\gamma}) =
F_{\gamma} A (1-\lambda)$$ photon events on average, where $A$ is the
exposure of the experiment for a given dataset and $\lambda$ is
the fraction of ``lost'' photons
\footnote{Due to fluctuations, a minor fraction
$\lambda$ of photons with $E>E_{\rm min}$ would have $E_{\rm
est}<10^{18}$~eV; we account for these ``lost photons'' as described in
Ref.~\cite{composition}.} (values of $\lambda $ are given in
Table~\ref{tab:results}). Let $\mathcal P(n)$ be the probability to have
$n$ photons in a dataset (calculated from data following
 Ref.~\cite{composition}). To constrain the flux $F_{\gamma}$ at the
confidence level $\xi$ one requires
$$
\sum_{n} \mathcal P(n) W(n,\bar n(F_{\gamma})) < 1 - \xi\,,
$$ where $W(n,\bar n)$ is the Poisson probability to observe $n$ particles
for the average $\bar n$.

\subsection{Results}
\begin{figure}
  \centering
\includegraphics[width=0.95\columnwidth]{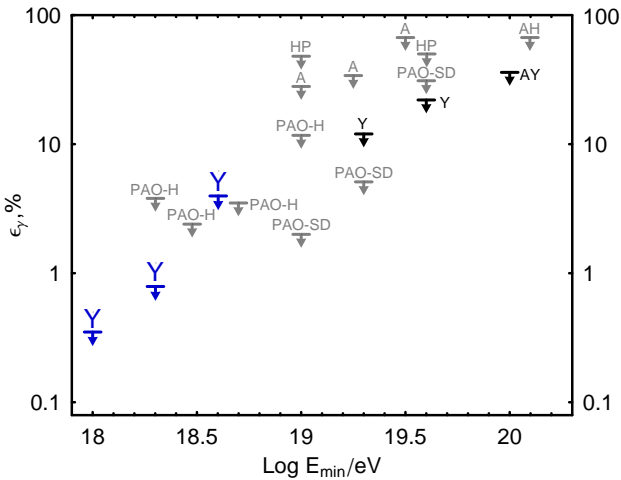}
\caption{
\label{fig:fraction}
Limits (95\% CL) on the fraction of primary gamma rays in the integral flux
of cosmic particles with $E_0>E_{\rm min}$ from:
this work (large Y);
hybrid events of the Pierre Auger Observatory (PAO-H)~\cite{PF};
the surface detector of the Pierre Auger Observatory (PAO-SD)
\cite{PAOsurface};
Yakutsk (small Y)~\cite{Ygamma};
reanalysis  of the AGASA (AH)~\cite{Homola} and AGASA and
Yakutsk (AY)~\cite{AYgamma} data; AGASA (A) \cite{AGASAmu} and Haverah
Park
(HP)~\cite{Haverah}.
}
\end{figure}

\begin{figure}
  \centering
\includegraphics[width=0.95\columnwidth]{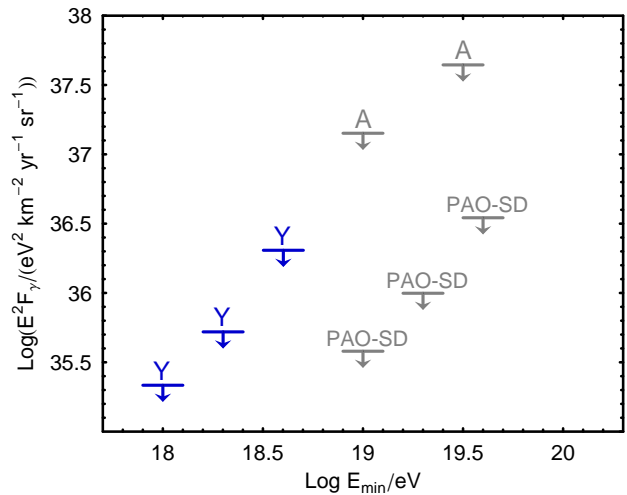}
\caption{
\label{fig:flux}
Limits (95\% CL) on the integral flux of primary gamma rays
with $E_0>E_{\rm min}$ from:
this work (Y);
the
surface detector of the Pierre Auger Observatory (PAO-SD)
\cite{PAOsurface} and
AGASA (A; assume mixed proton-gamma composition)
\cite{AGASAmu}.
}
\end{figure}

The upper limits on the observed flux and fraction of primary gamma rays
are summarized in Table~\ref{tab:results}.
\begin{table}
\begin{center}
\begin{tabular}{|c|c|c|c|}
\hline
$E_{\rm min}$, eV &  $10^{18}$ &  $2\times 10^{18}$ &  $4\times 10^{18}$ \\
\hline
$n_\gamma$ & 5.1 & 3.1 & 3.0\\
$F_\gamma$,  km$^{-2}$sr$^{-1}$yr$^{-1}$ & 0.22 & 0.13 & 0.13 \\
$E^2 F_\gamma$, $10^{35}$~eV$^2$km$^{-2}$sr$^{-1}$yr$^{-1}$ & 2.2 &
5.2& 20.8 \\
$\epsilon_\gamma$ & 0.004 & 0.008 & 0.041 \\
$\epsilon_\gamma$ ($E_{\rm est}+30\%$)&0.003 &0.005 &0.022 \\
$\epsilon_\gamma$ ($E_{\rm est}-30\%$)&0.006 &0.018 &0.108 \\
$N(E_{{\rm est}}>E_{\rm min})$ & 1647 & 341 & 63\\
$\lambda $ & 0.02 & $<0.01$ & $<0.01$\\
max$(p_{\gamma}^{(+)})$ & 0.25 & 0.026 & $<0.001$\\
$F_\gamma$,  km$^{-2}$sr$^{-1}$yr$^{-1}$, method~\cite{PF}   &0.25 &0.25
&0.25\\
\hline
\end{tabular}
\end{center}
\caption{
\label{tab:results}
Upper limits (95\% C.L.) on the number $n_\gamma$ of photons with
$E>E_{\rm min}$ in the sample, on the integral flux $F_\gamma$ of
photons with $E>E_{\rm min}$ and on the fraction $\epsilon_\gamma$ of
photons in the total integral flux of cosmic particles with $E>E_{\rm
min}$. The flux limits do not depend on the energy reconstruction
procedure; the fraction limits are given for the assumption of correct
energy reconstruction for non-photon primaries and for the supposed
overall shifts of $\pm30\%$ {\it for non-photon primaries}. Also given
are the number $N$ of events with $E_{\rm est} > E_{\rm min}$ in the
data set, the fraction of lost photons $\lambda $, the maximal
$p_{\gamma}^{(+)}$ for a given $E_{\rm min}$ and the limits on $F_{\gamma
}$ obtained by the statistical method of Ref.~\cite{PF} with our data.}
\end{table}
We compare the limits with those from previous works
in Fig.~\ref{fig:fraction} (for the gamma-ray
fraction)
and Fig.~\ref{fig:flux} (for the gamma-ray flux).

The energy range under study was partially explored by Auger
Collaboration, which placed a limit on the photon fraction at
$E>2\times 10^{18}$~eV~\cite{PF}.
We see that the result from
Yakutsk is stronger than Auger limits below $3\times 10^{18}$~eV, though
it is based on a smaller data set.
The principal reason
for this fact is related to the observable we use: the muon energy density
distinguishes primary photons from hadrons better than the depth of shower
maximum $X_{{\rm max}}$.

To illustrate this we applied the statistical method of Ref.~\cite{PF}
to our data. The Auger sample after selection and quality cuts contained
about 1050 events out of which 8 events were ``photon
candidates''~\cite{PF}. The latter were defined as events with
$X_{{\rm max}}$ exceeding those of photon initiated showers in 50\%
cases.  In Yakutsk we have 401 events of energy
$E>2\times 10^{18}$~eV \footnote{This number differs from
the one quoted in Table~\ref{tab:results} because of difference between
$E_{\rm est}$ and energy estimated under assumption of primary gamma
ray.}
with no photon candidates, that is $p_{\gamma}^{(+)} >0.5$ in our
language (maximal $p_{\gamma}^{(+)} =0.026$). These numbers
for various $E_{\rm min}$ together with limits on the gamma-ray
flux calculated by the statistical method of Ref.~\cite{PF} with our data
are also presented in Table~\ref{tab:results}. Our method gives slightly
more restrictive limits because the maximal  $p_{\gamma}^{(+)}$ is much
lower than the Auger threshold of 50\%.

The sensitivity of plastic scintillators to electromagnetic showers,
the strong discriminating power of large-area muon detectors, a 25-year
exposure and a sophisticated analysis
led up to the most stringent
limits on the primary photon flux at energies above $10^{18}$~eV and
$2\times 10^{18}$~eV. These limits
start to fill the gap between  limits on the
diffuse gamma-ray flux at $\lesssim \!\!10^{16}$~eV and $\gtrsim\!\! 10^{19}$~eV
and may
challenge previously allowed
new-physics models.
 The
{\it flux} limits do not depend on the energy reconstruction used by the
experiment (a reconstruction in assumption of primary photons is used),
nor on the simulations of hadronic showers. The {\it fraction} limits also
use the energy estimation in assumption of primary photons and also do not
rely on simulation of hadronic showers; however they depend on the assumed
energy estimation of non-photon primary particles~\footnote{Which uses
the attenuation curve determined from the constant intensity cuts and
the overall normalization to the measured air Cerenkov light and
therefore practically do not depend on simulations.}. This dependence is
weak in the high-statistics regime, cf.\ Table~\ref{tab:results}.

{\bf Acknowledgements. } We are indebted to L.~Dedenko,
O.~Kalashev and D.~Semikoz for helpful discussions.
This work was
supported in part by the RFBR
10-02-01406a and 09-07-00388a (INR team), 09-02-12028ofi-m (Yakutsk
team) and 08-02-00348a (MP), by the grants of the President of the
Russian Federation NS-5525.2010.2 (INR team), MK-61.2008.2 (GR) and
MK-1957.2008.2 (DG), by FASI
under state contracts 02.740.11.0244, 02.740.11.5092 (INR team), 02.740.11.0248,
02.518.11.7173 (Yakutsk team), by FAE under state contracts P520
and P2598 (INR team) and by the grant of the Dynasty Foundation (ST).
Numerical part of the work was performed at the cluster of the Theoretical
Division of INR RAS.


\end{document}